\documentstyle[12pt]{article}

\def\binom#1#2{{#1 \choose #2}}

\begin{document}

\author{Yi Cheng and Lin Zhang}
\title{$V$-Algebras and Their Free Field Realizations}
\date{}
\maketitle
\begin{center}
Department of Mathematics\\
University of Science and Technology of China\\
Hefei, Anhui 230026, P.R.China\\
E-mail: chengy@ustc.edu.cn, lzhang@nsc.ustc.edu.cn
\end{center}
\vspace{2cm}

\begin{center}
{\bf Abstract}
\end{center}

The $V$-algebras are the non-local matrix generalization of the well-known $W
$-algebras. Their classical realizations are given by the second Poisson
brackets associated with the matrix pseudodifferential operators. In this
paper, by using the general Miura transformation, we give the decomposition
theorems for the second Poisson brackets, from which we are able to
construct the free field realizations for a class of $V$-algebras including $%
V_{(2k,2)}$-algebras that corresponds to the Lie algebra of $C_k$-type as
the particular examples. The reduction of our discussion to the scalar case
provides the similar result for the $W_{\rm BKP}$-algebra.
\newpage

Recently, Bilal [1-3] has proposed non-local matrix generalizations of $W$%
-algebras, called $V_{(m,n)}$-algebras. Their classical realizations appears
naturally as the second Hamiltonian structures associated with the matrix
version of the $m^{\mbox{th}}$-order differential operators 
\begin{equation}
\label{1}L=\partial ^m+\sum_{j=0}^{m-2}U_j\partial ^j
\end{equation}
where $U_j$ are $n\times n$ matrix valued functions. The simplest example is
the $V_{(2,2)}$-algebra. It arises in the study of the non-abelian Toda
field theory as a model for strings propagating on a black hole background
[1] and its Poisson bracket version can be given by the second Hamiltonian
structure associated with [2] 
\begin{equation}
\label{2}L=\partial ^2-U,\ U=\left( 
\begin{array}{cc}
T & -
\sqrt{2}V^{+} \\ -\sqrt{2}V^{-} & T
\end{array}
\right) 
\end{equation}
The free field realization of the $V_{(2,2)}$-algebra was constructed first
by the factorization 
\begin{equation}
\label{3}L=\partial ^2-U=(\partial +P)(\partial -P)
\end{equation}
such that 
\begin{equation}
\label{4}U=P^{\prime }+P^2
\end{equation}
and then by expressing $P$ in terms of the vertex-like operators. In analogy
to the scalar case, (4) is called the Miura transformation.

In general, the Miura transformation can be given by the similar
factorization [3] 
\begin{equation}
\label{5}L=\partial ^m+\sum_{j=0}^{m-2}U_j\partial ^j=\prod_{j=1}^m(\partial
+P_j)
\end{equation}
and relates the second Poisson bracket of $U_j$ to much simpler ones of $P_j$%
, where $P_j$ are $n\times n$ matrices and satisfy the constraint $\sum P_j=0
$. However to the contrary of the scalar (i.e. $n=1$) case, the $P_j$ are not free
fields in general, and as far as we know, except $V_{(2,2)}$-algebras, it is
not clear how to give the free field realization for general $V_{(m,n)}$%
-algebra. As pointed by Bilal [3], the reason of this difference between the 
$W$-algebras and their matrix generalization $V$-algebras is the existence
of non-local terms in the $V$-algebras.

In this paper, we first generalize Bilal's $V_{(m,n)}$-algebra to the $V$%
-algebra associated with the $m^{\mbox{th}}$-order matrix pseudodifferential
operator (matrix $\Psi $DO) 
\begin{equation}
\label{6}L=\partial ^m+\sum_{j=-\infty }^{m-1}U_j\partial ^j
\end{equation}
where $U_j$ are $n\times n$ matrix valued functions, then give a general
decomposition theorem for the second Poisson bracket associated with (6) by
using the factorization $L=L_1L_2$, where both $L_1$ and $L_2$ are matrix $%
\Psi $DOs with the order being $m_1$ and $m_2$ respectively and satisfying $%
m_1+m_2=m$. It is not difficult to generalize the factorization to the
rational form $L=L_1L_2^{-1}$ since we may think that $L_1$ is factorized by 
$L_1=LL_2$. The above discussion is nothing but matrix generalization of our
previous work on the scalar case [4,5]. Thirdly, we consider the free field
realization of a more general class of $V$-algebras that correspond to the $L
$ of (6) with $2\times 2$ matrix coefficients and satisfying $L=L^{*}$ for a
proper defined adjoint action of the matrix $\Psi $DOs. Finally, as a
consequence, when we restrict to the scalar case, we obtain the $W$-algebras
represented by the second Poisson brackets of the BKP hierarchy and their
free field realization.

Let $L$ in (6) be the $m^{\mbox{th}}$-order matrix $\Psi $DO. For any
functional%
$$
\widetilde{f}=\int f(U_{m-1},U_{m-2},\cdots )dx 
$$
we define 
\begin{equation}
\label{7}\frac{\delta f}{\delta L}=\sum_{j=-\infty }^{m-1}\partial ^{-j-1}%
\frac{\delta f}{\delta U_j}
\end{equation}
and 
\begin{equation}
\label{8}(\frac{\delta f}{\delta U_j})_{\alpha \beta }=\sum_{r=0}^\infty
(-1)^r\frac{\partial ^r}{\partial x^r}(\frac{\partial f}{\partial
(U_j^{(r)})_{\beta \alpha }})
\end{equation}
is the matrix version of the Euler variation, where $(U_j^{(r)})_{\beta
\alpha }$ denotes the $(\beta ,\alpha )$ matrix element of $r^{\mbox{th}}$
derivative of $U_j$. Using (7) we find 
\begin{equation}
\label{9}d\widetilde{f}=<\frac{\delta f}{\delta L},\delta L>
\end{equation}
where the product $<\cdot ,\cdot >$ is defined by 
\begin{equation}
\label{10}<A,X>=\int \mbox{tr res}AXdx
\end{equation}
for any two matrix $\Psi $DOs of the form $A=\sum_{j=-\infty
}^{m-1}A_j\partial ^j$ and $X=\sum_{j=-\infty }^{m-1}\partial ^{-j-1}X_j$.
As in the scalar case, $A$ corresponds to the ``vector field'' $\partial _A$
and $X$ is called ``one form'' paired by (10) with the vector field. The
residues in (10) is defined to be the coefficient of $\partial ^{-1}$ term.

According to [3], The second Poisson bracket associated with (6) can be
defined in analogy with the scalar case [6] 
\begin{equation}
\label{11}\{\widetilde{f},\widetilde{g}\}_L=<H(\frac{\delta f}{\delta L}),%
\frac{\delta g}{\delta L}>
\end{equation}
where 
\begin{equation}
\label{12}H(X)=(LX)_{+}L-L(XL)_{+}=L(XL)_{-}-(LX)_{-}L
\end{equation}
mapping an one form $X=\sum_{j=-\infty }^{m-1}\partial ^{-j-1}X_j$ to the
vector field $\partial _{H(X)}$, where the subscripts ``$\pm $'' are
understood as the pure differential part or the residual part of the $\Psi $%
DO. If $U_{m-1}=0$ is assumed the following condition 
\begin{equation}
\label{13}\mbox{res}[\frac{\delta f}{\delta L},L]=0
\end{equation}
must be taken into account such that the leading coefficient of $\frac{%
\delta f}{\delta L}$ is expressed in terms of others. The second Poisson
bracket (11) constrained to $U_{m-1}=0$ is called the $V$-algebra [1-3].

The bracket (11) is bilinear and anti-symmetric because of the apparent
properties of the product (10). It will follow from the results on the Miura
transformation that for a class of matrix $\Psi $DOs used in this paper,
(11) also obeys the Jacobi identity. Nevertheless the following theorem does
not depend on the property of the Jacobi identity.

{\bf Theorem 1} {\it By the factorization} 
\begin{equation}
\label{14}L=L_1L_2
\end{equation}
{\it where} 
\begin{equation}
\label{15}L_i=\partial ^{m_i}+\sum_{j=-\infty }^{m_i-1}U_{ij}\partial ^j,\
i=1,2
\end{equation}
{\it are }$m_i^{\mbox{th}}${\it -order matrix $\Psi $DOs with }$m_1+m_2=m$%
{\it , then the Poisson bracket associated with }$L${\it \ of (6) is
decomposed to the summation of two brackets that are associated with }$L_1$%
{\it \ and }$L_2${\it \ respectively} 
\begin{equation}
\label{16}\{\widetilde{f},\widetilde{g}\}_L=\{\widetilde{f},\widetilde{g}%
\}_{L_1}+\{\widetilde{f},\widetilde{g}\}_{L_2}
\end{equation}
{\it If }$U_{m-1}=U_{m_1-1}+U_{m_2-1}=0${\it \ is assumed, then (13) is
equivalent to } 
\begin{equation}
\label{17}\mbox{res}[\frac{\delta f}{\delta L_1},L_1]+\mbox{res}[\frac{%
\delta f}{\delta L_2},\delta L_2]=0
\end{equation}

The proof of this theorem is essentially the same as we shown for the scalar
case in [4.,5], i.e. by (14) any functional $\widetilde{f}$ of $U_j$ is also
a functional of $U_{1j}$ and $U_{2j}$, therefore on the one hand we have%
$$
d\widetilde{f}=\int \mbox{tr res}\frac{\delta f}{\delta L}\delta Ldx=\int 
\mbox{tr res}\frac{\delta f}{\delta L}(\delta L_1L_2+L_1\delta L_2)dx 
$$
and on the other hand%
$$
d\widetilde{f}=\int \mbox{tr res}(\frac{\delta f}{\delta L_1}\delta L_1+%
\frac{\delta f}{\delta L_2}\delta L_2)dx 
$$
The above two expression imply that 
\begin{equation}
\label{18}\frac{\delta f}{\delta L_1}=L_2\frac{\delta f}{\delta L},\ \frac{%
\delta f}{\delta L_2}=\frac{\delta f}{\delta L}L_1
\end{equation}
each of them modular an $(-m_1-1)^{\mbox{th}}$-order and $(-m_2-1)^{\mbox{%
th}}$-order matrix $\Psi $DO respectively. Substitute (18) to the right hand
side of (16) and by the same calculation as that in [4,5], we can prove the
theorem.

It is easy to generalize Theorem 1 to the factorization $L=L_1\cdots L_r$,
in particular if $r=m$ and $L_j=\partial +P_j$, we immediately recover the
result of Bilal [3] 
\begin{equation}
\label{19}\{\widetilde{f},\widetilde{g}\}_L=\sum_{j=1}^m\int \mbox{tr res}[%
\frac{\delta f}{\delta P_j},\partial +P_j]\frac{\delta g}{\delta P_j}dx
\end{equation}
since the second Poisson bracket associated with $L_j=\partial +P_j$ is
simply 
\begin{equation}
\label{20}\{\widetilde{f},\widetilde{g}\}_{P_j}=\int \mbox{tr res}[\frac{%
\delta f}{\delta P_j},\partial +P_j]\frac{\delta g}{\delta P_j}dx
\end{equation}
The constraint $U_{m-1}=\sum P_j=0$ is then equivalent to 
\begin{equation}
\label{21}\sum_{j=1}^m(\frac{\delta f}{\delta P_j})^{\prime }=0
\end{equation}

{\bf Theorem 2} {\it If} 
\begin{equation}
\label{22}L=L_1L_2^{-1} 
\end{equation}
{\it where for the simplicity we assume that }$L_1${\it \ and }$L_2${\it \
are }$(m+k)^{\mbox{th}}${\it -order and }$k^{\mbox{th}}${\it -order matrix
differential operators respectively, then} 
\begin{equation}
\label{23}\{\widetilde{f},\widetilde{g}\}_L=\{\widetilde{f},\widetilde{g}%
\}_{L_1}-\{\widetilde{f},\widetilde{g}\}_{L_2} 
\end{equation}

The proof of this theorem can be completed simply by considering that $%
L_1=LL_2$ is factorized and then by applying Theorem 1. The scalar version
of the factorization $L=L_1L_2^{-1}$ was appeared in [7,8] for the study of $%
W$-algebras.

In the following we discuss the reduction of the second Poisson bracket (11)
to the subspace of matrix $\Psi $DOs that satisfy $L=L^{*}$. For the matrix $%
\Psi $DOs $A=\sum A_j\partial ^j$, we define the matrix version of the
adjoint action on $A$ by 
\begin{equation}
\label{24}A^{*}=\sum (-\partial )^j\sigma A_j^T\sigma ^{-1}
\end{equation}
where ``$T$'' denotes the matrix transposition, $\sigma $ is an $n\times n$
constant matrix such that the adjoint action satisfies 
\begin{equation}
\label{25}
\begin{array}{c}
(A^{*})^{*}=A\\
(AB)^{*}=B^{*}A^{*} \\ 
(A^{*})_{+}=(A_{+})^{*} \\ 
\int \mbox{tr res}A^{*}dx=-\int \mbox{tr res}Adx
\end{array}
\end{equation}
It is easy to see that such a matrix can be chosen freely as long as $%
\sigma $ is symmetric.

Let 
\begin{equation}
\label{26}W=L-L^{*}=\sum W_j\partial ^j
\end{equation}
where 
\begin{equation}
\label{27}W_j=U_j-\sum_{i=j}^{m-1}(-1)^i\binom i{i-j}\sigma \frac{\partial
^{i-j}U_i^T}{\partial x^{i-j}}\sigma ^{-1}
\end{equation}
then we can calculate that 
\begin{equation}
\label{28}\frac{\delta (W_j)_{\alpha \beta }}{\delta L}=(E_{\beta \alpha
}-(-1)^j\sigma E_{\alpha \beta }\sigma ^{-1} )\partial ^{-j-1}
\end{equation}
are symmetric 
\begin{equation}
\label{29}\frac{\delta (W_j)_{\alpha \beta }}{\delta L}=(\frac{\delta
(W_j)_{\alpha \beta }}{\delta L})^{*}
\end{equation}
with respect to the matrix version of adjoint action, where $E_{\alpha \beta
}$ are the $n\times n$ matrices only with the $(\alpha ,\beta )^{\mbox{th}}$
matrix element being equal to one and others to zero.

If we suppose that $m=2k$ and $L$ is symmetric 
\begin{equation}
\label{30}L=L^{*}
\end{equation}
(i.e. $W_j=0$), then the ``vector fields'' $\partial _A$ on the submanifold $%
W_j=0$ will be parametrized by the deformations of $L$ that remain
symmetric. These $A$ are clearly the matrix $\Psi $DOs of order at most $2k-1
$ obeying the symmetric property $A=A^{*}$. The ``one forms'' $X=\sum
\partial ^{-j-1}X_j$ on the submanifold $W_j=0$ must be chosen to be those
which are mapped via the Hamiltonian map $H$ to the vector fields $\partial
_{H(X)}$ tangent to the submanifold $W_j=0$. In other words, $(H(X))^{*}=H(X)$%
. Since 
\begin{equation}
\label{31}(H(X))^{*}=-H(X^{*}),
\end{equation}
$X$ must be anti-symmetric $X=-X^{*}$ modular a $(-m-1)^{\mbox{th}}$
order of matrix $\Psi $DO (i.e. the kernel of $H$). It can easily be checked
that these one forms are nondegenerately paired with the vector fields $%
\partial _A$, $A=A^{*}$. Actually we have checked that for some simple cases
for any functional $\widetilde{f}=\int fdx$ restricted on $W_j=0$, $X=\frac{%
\delta f}{\delta L}$ really satisfies $X=-X^{*}$ modular the kernel of $H$.

Therefore the Poisson bracket of two functionals $\widetilde{f}=\int fdx$
and $\widetilde{g}=\int gdx$ on the submanifold can be given by 
\begin{equation}
\label{32}
\begin{array}{c}
\{ 
\widetilde{f},\widetilde{g}\}_L=\frac 14<H(\frac{\delta f}{\delta L}-(\frac{%
\delta f}{\delta L})^{*}),\frac{\delta g}{\delta L}-(\frac{\delta g}{\delta L%
})^{*}> \\ =\frac 12<H(\frac{\delta f}{\delta L}-(\frac{\delta f}{\delta L}%
)^{*}),\frac{\delta g}{\delta L}> 
\end{array}
\end{equation}
with $L$ being symmetric.

The above argument is an analogue of that for the supersymmetric BKP
hierarchy [9]. The following theorem will provide another argument.

{\bf Theorem 3.} {\it If the }$m^{\mbox{th}}${\it -order (}$m=2k${\it )
symmetric matrix $\Psi $DO\ }$L${\it \ is factorized by} 
\begin{equation}
\label{33}L=L_1^{*}L_1
\end{equation}
{\it with} 
\begin{equation}
\label{34}L_1=\partial ^k+\sum_{j=-\infty }^{k-1}V_j\partial ^j
\end{equation}
{\it then we have} 
\begin{equation}
\label{35}\{\widetilde{f},\widetilde{g}\}_L=\frac 12\{\widetilde{f},%
\widetilde{g}\}_{L_1}
\end{equation}
{\bf Proof}: Any functional of $U_{m-1},U_{m-2},\cdots $ is also a
functional of $V_{k-1},V_{k-2},\cdots $ via the relation of (33). Therefore 
\begin{equation}
\label{36}d\widetilde{f}=<\frac{\delta f}{\delta L},\delta L>=<\frac{\delta f%
}{\delta L},\delta L_1^{*}L_1+L_1^{*}\delta L_1>
\end{equation}
$$
d\widetilde{f}=<\frac{\delta f}{\delta L_1},\delta L_1> 
$$
so 
\begin{equation}
\label{37}\frac{\delta f}{\delta L_1}=((\frac{\delta f}{\delta L})-(\frac{%
\delta f}{\delta L})^{*})L_1^{*}
\end{equation}
modular a $(-k-1)^{\mbox{th}}$-order matrix $\Psi $DO.

Substitute this expression to the Poisson bracket $\{\cdot ,\cdot \}_{L_1}$
with respect to $L_1$ we have the Poisson bracket (32) with respect to $L$,
which can be expressed by (35).

We may continue to factorize $L_1$%
\begin{equation}
\label{38}L_1=\prod_{j=1}^l(\partial +P_j)^{-1}\prod_{j=l+1}^{k+2l}(\partial
+P_j)
\end{equation}
where $l$ is an arbitrary integer, and then apply Theorem 2 and 3, we find
that the Poisson bracket (32) of $L=L_1^{*}L_1$ becomes 
\begin{equation}
\label{39}\{\widetilde{f},\widetilde{g}\}_L=\frac 12\sum_{j=l+1}^{k+2l}\{%
\widetilde{f},\widetilde{g}\}_{P_j}-\frac 12\sum_{j=1}^l\{\widetilde{f},%
\widetilde{g}\}_{P_j}
\end{equation}
with each $\{\widetilde{f},\widetilde{g}\}_{P_j}$ being given by (20).

Let us now calculate the coefficient of the second leading term of $%
L=L_1^{*}L_1$ with $L_1$ being in (38). It is 
\begin{equation}
\label{40}U_{m-1}=(-1)^k\sum_{j=l+1}^{k+2l}(P_j-\sigma P_j^T\sigma
^{-1})-(-1)^l\sum_{j=1}^l(P_j-\sigma P_j^T\sigma ^{-1})
\end{equation}
We immediately find that a sufficient condition of $U_{m-1}=0$ is 
\begin{equation}
\label{41}P_j-\sigma P_j^T\sigma ^{-1}=0,\ j=1,2,\cdots ,k+2l 
\end{equation}
namely the restriction of the Poisson bracket (32) of $L$ to the submanifold 
$U_{m-1}=0$ can be realized if each copy of the Poisson bracket in the form
of (20) associated with $\partial +P_j$ can be restricted to the submanifold
of (41).

According to the above analysis, we choose $n=2$, 
\begin{equation}
\label{42}\sigma =\left( 
\begin{array}{cc}
0 & 1 \\ 
1 & 0
\end{array}
\right) 
\end{equation}
and $k+2l$ copies of Bilal's $V_{(2,2)}$-algebra 
\begin{equation}
\label{43}P_j=\left( 
\begin{array}{cc}
T_j & -
\sqrt{2}V_j^{+} \\ -\sqrt{2}V_j^{-} & T_j
\end{array}
\right) ,\ 1\leq j\leq k+2l
\end{equation}
among them the first $l$ copies have a sign difference with the $V_{(2,2)}$%
-algebra. It is obvious that $P_j$ obey (41) and their elements can be
expressed in terms of $k+2l$ independent groups of vertex-like fields. Thus
we may construct the free field realization of the $V$-algebra that
corresponds to the second Poisson bracket on the space of $m^{\mbox{th}}$%
-order ($m=2k$) and $2\times 2$ matrix $\Psi $DOs restricted by $L=L^{*}$. A
simple case is for $l=0$, i.e. if $L_1$ is a pure differential operator, so 
\begin{equation}
\label{44}L=L_1^{*}L_1=(-1)^k(\partial -P_k)\cdots (\partial -P_1)(\partial
+P_1)(\partial +P_k)
\end{equation}
the $V_{(2k,2)}$-algebra in this case corresponds to the Lie algebra of the $%
C_k$-type [3]. Our result give its free field realization. Note that from
mathematical point of view, the $(-1)^k$ factor does not affect the
structure of Poisson bracket essentially.

Finally we are going to restrict the above results to the scalar case $n=1$
and connect the Poisson bracket (32) for $n=1$ with the Poisson bracket for
the BKP hierarchy. Let 
\begin{equation}
\label{45}\Lambda =\partial ^{2k+1}+\sum_{j=-\infty }^{2k}v_j\partial ^j
\end{equation}
where $v_j$ are scalar functions. Then we define 
\begin{equation}
\label{46}L=\partial \Lambda =\partial ^{2k+2}+\sum_{j=-\infty
}^{2k+1}u_j\partial ^j
\end{equation}
The relation between $v_j$ and $u_j$ can be given explicitly 
\begin{equation}
\label{47}
\begin{array}{c}
u_{2k+1}=v_{2k} \\ 
u_j=v_j^{\prime }+v_{j-1}
\end{array}
\end{equation}
from which we first have%
$$
\frac{\partial f}{\partial v_j^{(l)}}=\frac{\partial f}{\partial
u_{j+1}^{(l)}}+\frac{\partial f}{\partial u_j^{(l-1)}} 
$$
and so%
$$
\frac{\delta f}{\delta v_j}=\frac{\delta f}{\delta u_{j+1}}-(\frac{%
\delta f}{\delta u_j})^{\prime } 
$$
which implies that 
\begin{equation}
\label{48}\frac{\delta f}{\delta L}=\frac{\delta f}{\delta \Lambda }\partial
^{-1}+(\mbox{a }(-2k-3)^{\mbox{th}}\mbox{-order }\Psi DO\mbox{)}
\end{equation}
If we assume that $\Lambda $ is the $\Psi $DO associated with the BKP
hierarchy [10] 
\begin{equation}
\label{49}\Lambda ^{*}=-\partial \Lambda \partial ^{-1}
\end{equation}
then $L$ in (46) is symmetric $L=L^{*}$ and $u_{2k+1}=v_{2k}=0$ where the
adjoint action on the scalar $\Psi $DO $A=\sum a_j\partial ^j$ is defined as
usual $A^{*}=\sum (-\partial )^ja_j$. Substitute (48) into (32) for $n=1$ we
notice that the second term of the right hand side does mot contribute
anything and the Poisson bracket in terms of $\Lambda $ is given by 
\begin{equation}
\label{50}
\begin{array}{c}
\{\widetilde{f},\widetilde{g}\}_\Lambda =\int \mbox{res}[\partial
^{-1}(\partial \Lambda \frac{\delta f}{\delta \Lambda }\partial
^{-1})_{+}\partial \Lambda -\Lambda (\frac{\delta f}{\delta \Lambda }\Lambda
)_{+}+\partial ^{-1}(\partial \Lambda \partial ^{-1}(\frac{\delta f}{\delta
\Lambda })_{+}^{*}\partial \Lambda  \\ -\Lambda (\partial ^{-1}(\frac{\delta
f}{\delta \Lambda })^{*}\partial \Lambda )_{+}]\frac{\delta g}{\delta
\Lambda }dx
\end{array}
\end{equation}
We define $W_{\rm BKP}^{(2k+1)}$-algebra corresponding to the second
Poisson bracket (50) associated with the BKP hierarchy. Its free field
realization is given by the following factorization 
\begin{equation}
\label{51}\Lambda =\partial ^{-1}L_1^{*}L_1
\end{equation}
with 
\begin{equation}
\label{52}L_1=\prod_{j=1}^l(\partial
+p_j)^{-1}\prod_{j=1+1}^{k+1+2l}(\partial +p_j)
\end{equation}
where $p_j$ are independent fields and satisfy 
\begin{equation}
\label{53}
\begin{array}{c}
\{p_i(x),p_j(y)\}=-\delta _{ij}\delta ^{\prime }(x-y)\ 1\leq i,j\leq l \\ 
\{p_i(x),p_j(y)\}=\delta _{ij}\delta ^{\prime }(x-y)\ l+1\leq i,j\leq k+1+2l
\\ 
\{p_i(x),p_j(y)\}=0\ 1\leq i\leq l,\ l+1\leq j\leq k+1+2l
\end{array}
\end{equation}

In conclusion we have discussed the properties of the second Poisson
structure associated with the matrix $\Psi $DO. These properties enable us
to construct the free field realizations for a more general class of $V$%
-algebras that correspond to the second Poisson brackets of matrix $\Psi $%
DO. Because of the non-locality of the $V$-algebras, the free field
realizations for them become more difficult than for $W$-algebras. It would
be of interest to investigate these problems for the general $V$-algebras.

\section*{Acknowledgements}

This work was supported by the National Basic Research Project for
``Nonlinear Science'' and Fund of CAS.

\end{document}